\documentstyle[12pt]{article}
\input{epsf}

\begin{document}

\vspace*{3mm}

\begin{center}

{\LARGE \bf
The Higgs-Yukawa model in curved spacetime}

\vspace{6mm}

 {\bf E. Elizalde} \\
Center for Advanced Study CEAB, CSIC, 17300 Blanes
and Department ECM and IFAE, Faculty of Physics,
University of  Barcelona, Diagonal 647, 08028 Barcelona, Spain
\footnote{E-mail: eli@zeta.ecm.ub.es}\\ and  \\
{\bf S.D. Odintsov} \\
Department ECM, Faculty of Physics,
University of  Barcelona \\  Diagonal 647, 08028 Barcelona,
Spain\footnote{E-mail: odintsov@ecm.ub.es      } \\
and Tomsk Pedagogical Institute, 634041 Tomsk, Russia \\

\vspace{15mm}

{\bf Abstract}

\end{center}

The Higgs-Yukawa model in curved spacetime
 (renormalizable    in the usual sense)
 is considered near the critical  point, employing the
$1/N$--expansion
and renormalization group techniques. By making use of the equivalence
of this
 model with the standard NJL model,
the effective potential in the linear curvature approach
is calculated and the
 dynamically generated fermionic mass is found.
A numerical study of chiral symmetry breaking by curvature effects is
 presented.

\vspace{4mm}

\newpage


 The Nambu-Jona-Lasinio  (NJL) model \cite{1}
and the Gross-Neveu model \cite{2} belong to a very restricted class
of quantum theoretical models in which an analytical study of the
composite bound states is possible. Such models are usually studied
in the framework of the $1/N$--expansion (see \cite{3} for a review).

Recently, there has been some interest in the literature \cite{4,5} about the
dynamical symmetry breaking pattern of NJL-like models for the
electroweak interaction, where the top quark plays the role of
order parameter. A study of the NJL model in curved spacetime
has been carried out in Refs. \cite{6}-\cite{8}. Using a
block-spin renormalization group (RG), the existence of an
IR stable fixed point in the NJL model was pointed out \cite{6,4}.

In Ref. \cite{9} it was suggested that it would be interesting to
consider the Higgs-Yukawa (or simply Yukawa) model,
which is multiplicatively renormalizable in
the usual sense.
In frames of the $1/N$--expansion
the NJL model and the Higgs-Yukawa       model near the critical
point become completely equivalent and they describe
the same physics of chiral symmetry breaking (CSB).

In the present work we consider the   Higgs-Yukawa
    model in curved spacetime and make use of its
equivalence with
the NJL model in curved spacetime.
The effective
potential is found in the large-$N $     (and  linear
curvature)  approximation at finite
cut-off, and also after removing the cut-off in the coupling
 constants
(a la Coleman-Weinberg \cite{10}). The dynamical fermionic mass
in curved spacetime is calculated.


 We start now the study of the renormalizable
Higgs-Yukawa model                        in curved spacetime.
 We will be
interested in the dynamics of this theory as a kind of
four-fermion model near the critical point.

The (multiplicatively renormalizable) Lagrangian of the theory under
discussion is the typical one of a Yukawa-type interaction \cite{11}
\begin{eqnarray}
L&=& \frac{1}{2} g^{\mu\nu} \partial_\mu \sigma \partial_\nu \sigma -
 \frac{1}{2}
m^2 \sigma^2 +  \frac{1}{2} \xi R \sigma^2 -  \frac{\lambda }{4!} \sigma^4 +
 \bar{\psi} (i\gamma^\mu  (x) \nabla_\mu - h \sigma) \psi \nonumber \\
&&+ \Lambda + \kappa R + a_1 R^2 + a_2 C^2_{\mu\nu\alpha\beta} + a_3G + a_4
\Box R, \label{1}
\end{eqnarray}
where $\sigma$ is a scalar field,  $\psi$  a massless, $N$-component Dirac
spinor, $g^{\mu\nu}$  an arbitrary metric of the classical gravitational
 field and, for simplicity, we will set $\Box R =0$.
    Here only $\sigma$ and $\psi$ will be quantized.

The restricted version of  this theory  with
\begin{equation}
L= \bar{\psi} (i\gamma^\mu  (x) \nabla_\mu - h \sigma) \psi
- \frac{1}{2}m^2\sigma^2,  \label{2}
\end{equation}
where $\sigma$ is to be treated as an auxiliary scalar field, represents
(after elimination of $\sigma$) the standard four-fermion model  of NJL
type
\begin{equation}
L_{\mbox{ four-fermion}}=
 \bar{\psi} i\gamma^\mu  (x) \nabla_\mu    \psi
+ \frac{h^2}{m^2} (\bar{\psi} \psi)^2,  \label{3}
\end{equation}
where it is convenient to put $h=1$ and $m^2 = N/(2\lambda_1)$.

The Yukawa model                      of the kind (\ref{1})
near the critical point
 in flat space has been considered  in Ref.
\cite {9}, where it was pointed out that the physics of that model is
again
 the
physics of CSB. Under a chiral transformation, $\sigma$ transforms into
$-\sigma$.

We will now study in detail the Higgs-Yukawa       model (\ref{1}), by making
use of
 the fact that it is multiplicatively renormalizable. The standard one-loop
$\beta$-functions can be found easily
(see \cite{26} for flat space and \cite{11a} for curved space).

By direct inspection of the $\beta$-function,
 one can see in the matter sector the  fixed point $\lambda =g^2 =
m^2 =0$, $\xi =1/6$ is infrared stable. In the  IR limit, i.e. at a scale
$\mu  << \Lambda$, one finds for the dimensionless running couplings ($t=
\frac{1}{2} \ln (\mu^2 /\Lambda^2)$)
\begin{eqnarray}
&& h^2(t)  \sim - \frac{(4\pi)^2}{(4N+6)t}, \ \ \  \xi (t) \sim \frac{1}{6} +
 \left( \xi
-  \frac{1}{6} \right) t^{-\lambda_* - 4N/(4N +6)}, \nonumber \\
&& \lambda (t) \sim
 -\frac{(4\pi)^2}{t} \,  \lambda_*, \   \lambda_* = \frac{1}{6(2N +3)} \left[ -
(2N-3) + \sqrt{ 4N^2 + 132 N +9} \right], \label{6}
\end{eqnarray}
and from here one can easily get the corresponding relations for
large $N$. These relations give the logarithmic corrections to the IR
fixed point solution in the matter sector. Moreover, as one can see the
 behavior
of the couplings of the Higgs-Yukawa       model in curved spacetime
(in the matter sector) near the critical
point is the same  as for the usual non-renormalizable NJL model in curved
spacetime \cite{6}.

Using Eq. (\ref{6}) we can investigate the behavior of the
scalar-graviton coupling constant $\xi$ for the composite bound state.
Choosing (for an estimation) $\Lambda \simeq M_{Pl} \simeq 10^{19}$ GeV
and $\mu \simeq M_{GUT} \simeq 10^{15}$ GeV, at large $N$ we obtain
\begin{equation}
\xi (t) \sim  \frac{1}{6} -\frac{1}{8} \left( \xi - \frac{1}{6}\right),
 \label{e8}
\end{equation}
i.e. being the IR limit
 an asymptotically conformal one, as in \cite{6} (see also \cite{14},
and for GUTs \cite{12}), our $\xi (t)$
 still
depends  ---weakly---  on the initial value $\xi= \xi (0)$. For the choice $\xi
 =0$,
 $\xi
 (t)$ becomes positive and not large,  what may be relevant for  cosmological
applications where, for instance, the model of extended inflation \cite{15}
 favors
very small negative values of $\xi$  (corresponding to $\xi \simeq 4/3$ in
 (\ref{e8})),
while the inflationary model  of Refs. \cite{16} favors very large and negative
values of $\xi$.


 Let us now proceed
 with
 the study of the effective potential for composite fields in curved spacetime.
Rewritting the matter sector of the Lagrangian as follows
\begin{equation}
L= \frac{1}{2h^2} g^{\mu\nu} \partial_\mu \sigma \partial_\nu \sigma
-\frac{N}{2\lambda_1 h^2} \sigma^2 +\frac{N}{2\xi_1 h^2} R \sigma^2
 -\frac{N}{4! \lambda_2 h^4} \sigma^4 +
\bar{\psi} (i\gamma^\mu  (x) \nabla_\mu -   \sigma) \psi , \label{8}
\end{equation}
where we have performed the substitutions  $m^2 \rightarrow N/(2\lambda_1)$,
  $\xi \rightarrow N/(2\xi_1)$ and  $\lambda \rightarrow N/\lambda_2$, and
$\sigma$ has been rescaled into $h\sigma$, for simplicity, and if we suppose
that $\sigma$ is a constant, we can work  out the semiclassical effective
action
$S_{eff}$. Integrating over the fermion fields, we obtain
\begin{equation}
S_{eff} =\int d^4x \, \sqrt{-g} \, \left(
-\frac{1}{2\lambda_1 h^2} \sigma^2 +\frac{1}{2\xi_1 h^2} R \sigma^2
 -\frac{1}{4! \lambda_2 h^4} \sigma^4 \right) -i \ln \det
\left(i\gamma^\mu  (x) \nabla_\mu -   \sigma \right), \label{9}
\end{equation}
 where $N$ has been factored out.

Now, working in the linear curvature approximation and regularizing the
 divergent integrals by the cut-off method (see, for example,
Ref. \cite{25}), we can repeat the calculations of Refs.
 \cite{7,8}
to obtain, for our model,
\begin{eqnarray}
V(\sigma) &=& V(0)+ \frac{1}{2\lambda_1 h^2} \sigma^2 -
\frac{1}{2\xi_1 h^2} R \sigma^2
 +\frac{1}{4! \lambda_2 h^4} \sigma^4 \nonumber \\
&&- \frac{1}{(4\pi)^2} \left[ \sigma^2\Lambda^2 + \Lambda^4  \ln \left( 1 +
 \frac{\sigma^2}{\Lambda^2} \right) -\sigma^4  \ln \left( 1 +
 \frac{\Lambda^2}{\sigma^2} \right)\right] \nonumber \\
&&+ \frac{R}{6(4\pi)^2} \left[  -\sigma^2  \ln \left( 1 +
 \frac{\Lambda^2}{\sigma^2} \right) +\frac{ \sigma^2\Lambda^2}{\sigma^2
+\Lambda^2}  \right] . \label{10}
\end{eqnarray}
We have  thus obtained the effective potential $V(\sigma)$
in the linear curvature approximation and for a finite cut-off.

We will here adopt a different strategy,  by making use of the crucial
property that the above model           is a multiplicatively
renormalizable
 theory.
 Owing to this fact, we  may repeat the analysis of Coleman and Weinberg
\cite{10}, and  throw away terms which vanish when $\Lambda^2$ goes to infinity
in (\ref{10}), to remove the  remaining $\Lambda^2$-dependent terms via the
renormalization of the coupling constants by imposing Coleman-Weinberg type
renormalization conditions (for a general description in the case of a curved
spacetime, see \cite{16a}). As a result, after some algebra we obtain
\begin{equation}
V(\sigma) =  \frac{1}{2\lambda_1 h^2} \sigma^2 -
\frac{1}{2\xi_1 h^2} R \sigma^2
 +\frac{1}{4! \lambda_2 h^4} \sigma^4 + \frac{\sigma^4}{(4\pi)^2} \left(
 \ln \frac{\sigma^2}{\mu^2}- \frac{25}{6}  \right) +
\frac{R\sigma^2}{6(4\pi)^2}
 \left( \ln \frac{\sigma^2}{\mu^2}- 3  \right) , \label{11}
\end{equation}
where $\mu^2$ is the mass parameter. Notice that using the form  (\ref{11}) of
 the effective potential it is  more difficult to compare the properties of the
 Higgs-Yukawa
    model with those of the usual NJL model, due to the fact that some cut-off
dependence is hidden in the new parameters $\xi_1$ and $\lambda_2$. (One
might also adopt another point of view  and hide the new parameters in
the cut-off procedure, as  was suggested in \cite{9}).

Let us now analyze the phase structure  of the effective potential (\ref{11}).
The dynamical mass of the fermion is calculated by using  the gap equation
\begin{equation}
\left. \frac{\partial V(\sigma )}{\partial \sigma} \right|_{\sigma = \sigma_0}
 = \frac{\sigma_0}{\lambda_1 h^2}  -
\frac{R\sigma_0}{\xi_1 h^2}
 +\frac{\sigma_0^3}{6  \lambda_2 h^4}
+ \frac{\sigma_0^3}{4\pi^2} \left(
\ln \frac{\sigma_0^2}{\mu^2}- \frac{11}{3}  \right)
 + \frac{R\sigma_0}{3(4\pi)^2} \left(
\ln \frac{\sigma_0^2}{\mu^2}- 2  \right) =0.  \label{12}
\end{equation}

The solution $\sigma_0$ of Eq. (\ref{12}) corresponds to the ground state of
the
composite field $\bar{\psi} \psi$ (near the critical point) and is equal to the
dynamical mass of the fermion. Supposing that such a solution exists, and
choosing  $\mu^2 = \sigma_0^2$, we get
\begin{equation}
\sigma_0^2 = \left(  \frac{R}{\xi_1 h^2} -\frac{ 1}{\lambda_1 h^2} +
\frac{2R}{3(4\pi)^2}\right) \left( \frac{1}{6\lambda_2 h^4}
 -\frac{11}{12(4\pi)^2}
\right)^{-1}. \label{13}
\end{equation}
As we can see,  in absence of  gravitational field,  chiral symmetry
breaking takes place at  $$ \frac{11}{2(4\pi)^2} > \frac{1}{\lambda_2 h^4}. $$
 On
the contrary, in the presence of a gravitational field there may arise a
 completely
gravitational effect, which can occur even in the situation when there is no
 CSB
in flat space.  Choosing different values for the coupling constants in the
effective potential (\ref{11}), one can numerically investigate the phase
 structure
of the theory, and in particular, the curvature-induced phase transitions
\cite{11,16a} between the CS phase and the CSB phase.

In Fig. 1 we show the potential (\ref{11}) for different values of $R$
and fixed values of $\lambda_1$, $\lambda_2$, $\xi_1$, $h$ and $\mu$.
The following combinations of adimensional variables have been chosen
(they appear naturally).
For the potential itself $f(x) \equiv
V(x) /\mu^4$ (this is the $y$-axis)
 ---being the variable $x\equiv \sigma^2/\mu^2$--- and for the
coefficients of the different terms $a_1 \equiv (2\lambda_1 h^2
 \mu^2)^{-1}$, $a_2\equiv (2\xi_1 h^2)^{-1}$, $a_3 \equiv (4! \,
\lambda_2 h^4)^{-1}$ and $r \equiv R/\mu^2$, what yields the function
\begin{equation}
f(x) = a_1x -a_2 r x+ a_3 x^2 + \frac{x^2}{(4\pi)^2} \left( \ln x -
25/6 \right) + \frac{r x}{6(4\pi)^2} \left( \ln x -3 \right).
\end{equation}
We have simply taken $a_1=a_2=a_3 =1$ and $r=0,3,10,20$, and the
 potential is compared in the same range $0 \leq x \leq 15$.
The upper curve corresponds to the zero curvature case
and the curvature increases as we
go down.
\begin{figure}
\vskip-3cm
\centerline{\epsfxsize=11cm \epsfbox{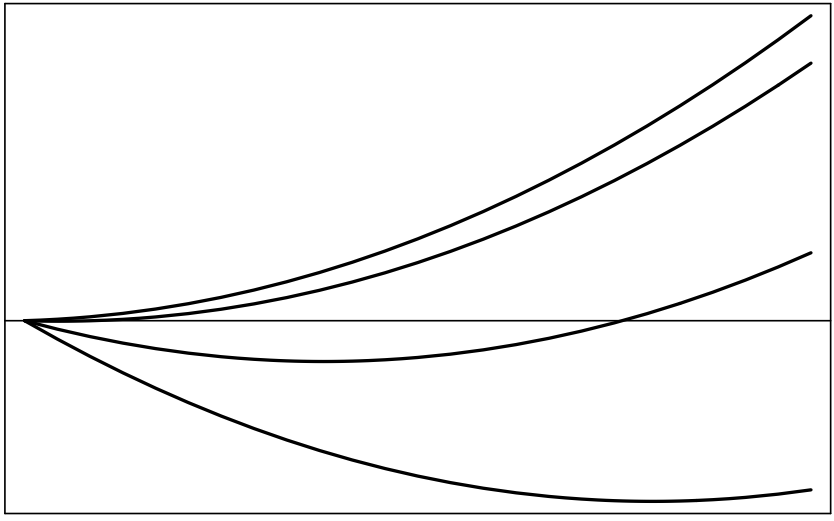}}
\vskip-35mm
\caption{{\protect\small
Plot of the potential $V(\sigma^2/\mu^2) /\mu^4$ in terms of $\sigma^2
/\mu^2$ for different values of $R$
and fixed values of $\lambda_1$, $\lambda_2$, $\xi_1$, $h$ and $\mu$.
The upper curve corresponds to $R=0$ and the curvature increases as we
go down.}}
\label{f1}
\end{figure}



It is interesting to remark that in the $1/N$ expansion one can consider
the gravitational field to be a quantum field as well, since this does
not change at all the picture.
 Working  in the
$1/N$--expansion (as was proposed some time ago \cite{18}), using  the
large-$N$ contributions to the $\beta$-functions, we get in the
 infrared
regime ($t \rightarrow -\infty$) Eqs. (\ref{6}) (dropping the next to
leading terms) together with the following equations for the
gravitational couplings:
 \begin{eqnarray}
 && \Lambda (t) \sim
 \Lambda - \frac{m^4}{2(4\pi)^2t}, \ \ \ \ \ \ \ \ \kappa (t) \sim
 \kappa - \frac{m^2(\xi -1/6)}{2(4\pi)^2 t}, \nonumber \\
&&\alpha (t) = \frac{\alpha}{1 + \displaystyle\frac{\alpha t}{20
(4\pi)^2}},  \ \ \ \
 \ \ \ \
 \eta^{-1} (t)
\simeq  -\frac{(\xi -1/6)^2}{2(4\pi)^2Nt},
\label{14}
\end{eqnarray}
where $\alpha =N/a_2$ and $\eta = N/a_1$.
Thus, we see that  the cosmological and Newtonian
couplings are
growing in the infrared regime, as compared with the matter couplings which
go to their IR fixed-point values. Notice that proper QG corrections to the
$\beta$-functions are negligible in the $1/N$--expansion.
 The
behavior of $\alpha (t)$  coincides  precisely with  the corresponding  one
 that
was obtained  in \cite{18}, where a theory of $N$ massless free fermions
interacting with QG was considered.  For $\alpha <0$ we get  asymptotic freedom
 for $\alpha (t)$ (as is the case for $\eta^{-1}$ in the IR regime).
Thus we are able to discuss the running of the effective couplings in
the quantum gravity-Higgs-Yukawa system in the infrared region.

We think that the
$1/N$--expansion, which is a gauge invariant procedure, provides very
interesting possibilities to deal with QG-matter systems.


  In summary, we have studied here the effective potential for the
Higgs-Yukawa model
      in curved spacetime near the critical point
where it is equivalent to the standard NJL model.  We have also
calculated
 the dynamical fermionic mass and studied the curvature-induced phase
transitions of the model.
We envisage
other possibilities to extend the NJL model while beeing still able to use  the
$1/N$--expansion. In particular, let us mention  the higher-derivative
NJL models \cite{20,21}, where it would be interesting to apply a similar
 analysis.

\bigskip

\noindent{\bf Acknowledgments}.

We thank C.T. Hill for correspondence and the referee for some positive
suggestions.
SDO would like to acknowledge the  hospitality of the
members of the Department ECM, Barcelona University. This work has been
supported  by  DGICYT
(Spain) and  by CIRIT (Generalitat de Catalunya).

\end{document}